\documentclass[a4paper, 11pt]{article}

\usepackage{amssymb}
\usepackage{latexsym}
\usepackage{amsmath}

\def\be{\begin{equation}}
\def\ee{\end{equation}}
\def\bea{\begin{eqnarray}}
\def\eea{\end{eqnarray}}

\def\nn{\nonumber}
\def\fft{\frac}

\begin{document}

  \begin{flushright}
    hep-th/0508169 \\
    DAMTP-2005-77
  \end{flushright}
  \vskip 1cm
  \begin{center}
    \LARGE
   {\bf Special symmetries of the charged Kerr-AdS black hole of
   $D=5$ minimal gauged supergravity}
  \end{center}
  \vskip 0.5cm
  \begin{center}
        \begin{center}
        {\Large Paul Davis\footnote{P.Davis@damtp.cam.ac.uk}, Hari
        K. Kunduri\footnote {H.K.Kunduri@damtp.cam.ac.uk} and James
        Lucietti\footnote{J.Lucietti@damtp.cam.ac.uk}  }\\
        \bigskip\medskip
        {\it  DAMTP, Centre for Mathematical Sciences,\\
      University of Cambridge,
      Wilberforce Rd.,\\
      Cambridge CB3 0WA, UK\\}
        \end{center}
  \end{center}
  \vskip 0.5cm
\begin{center}
{\bf Abstract}
\end{center}
\centerline{
\parbox[t]{15cm}{\small
\noindent In this note we prove that the Hamilton-Jacobi equation in
the background of the recently discovered charged Kerr-AdS black
hole of $D=5$ minimal gauged supergravity is separable, for
arbitrary values of the two rotation parameters. This allows us to
write down an irreducible  Killing tensor for the spacetime. As a
result we also show that the Klein-Gordon equation in this
background is separable. We also consider the Dirac equation in
this background in the special case of equal rotation parameters and
show it has separable solutions. Finally we discuss the near-horizon geometry
of the supersymmetric limit of the black hole.}}

\vskip1.2cm

\noindent It is a curious fact that the Kerr-Newman black hole
possesses a hidden symmetry which renders geodesic motion
integrable~\cite{Carter1}. This is related to the existence of a
second rank Killing tensor $K_{\mu\nu}$; by definition such a tensor
satisfies $\nabla_{(\mu} K_{\nu \rho)}=0$. Given a Killing tensor
one may construct the quantity $K=K_{\mu\nu} \dot{x}^{\mu}
\dot{x}^{\nu}$ which is conserved along geodesics $x^{\mu}(\tau)$.
Carter was the first to systematically analyse the consequence of
separability of solutions to Einstein's equations, and indeed this
is how the Kerr-(A)dS black hole and its charged counterpart were
first discovered~\cite{Carter2}. Higher dimensional Kerr-(A)dS
metrics have only been recently constructed~\cite{GLPP}. The
existence of a Killing tensor has been verified in five dimensions
for arbitrary rotation parameters~\cite{KL}, in all dimensions for
the special cases of equal sets of rotation
parameters~\cite{VSP,VS}. This renders both the
Hamilton-Jacobi (HJ) and Klein-Gordon (KG) equations separable. The
charged counterparts of the Kerr-AdS black holes in $D=5$ minimal
gauged supergravity are far more difficult to construct. Progress
was first made by tackling the special case where the rotation
parameters are equal~\cite{CLP}, and a reducible Killing tensor for
this black hole was found in~\cite{KL2}. Further, the same Killing
tensor was found for black holes with equal rotation parameters in
the more general $U(1)^3$ theory~\cite{vasu}. Only very recently has
a charged Kerr-AdS black hole been found with arbitrary rotation
parameters~\cite{CCLP}. The purpose of this note is to show that
this black hole also has a Killing tensor rendering geodesic motion
integrable and the KG equation separable. We also show that the
Dirac equation admits separable solutions in the special case of
equal rotation parameters. Finally we discuss the near-horizon
geometry of the supersymmetric limit of this black hole. In the case
of equal rotation parameters we show that it has a symmetry algebra
$\mathfrak{sl}(2,\mathbb{R})  \times \mathfrak{su}(2) \times
\mathfrak{u}(1)$ as is the case for the BMPV black hole~\cite{GMT}.

\noindent
In~\cite{CCLP} it was shown that $D=5$ minimal gauged supergravity
with the Lagrangian density
\be
{\cal L} = (R+ 12g^2)\, {* 1} - \frac{1}{2} \, {F}\wedge *F + \fft1{3\sqrt3}
      F\wedge F\wedge A\,
\ee
where $F=dA$, admits a black hole solution parameterised by its mass,
charge and two rotation parameters. Explicitly the metric is given by:
\bea
ds^2 &=& -\fft{\Delta_\theta\, [(1+g^2 r^2)\rho^2 d t + 2q \nu]
\, d t}{\Xi_a\, \Xi_b \, \rho^2} + \fft{2q\, \nu\omega}{\rho^2}
+ \fft{f}{\rho^4}\Big(\fft{\Delta_\theta \, d t}{\Xi_a\Xi_b} -
\omega\Big)^2 + \fft{\rho^2 d r^2}{\Delta_r} +
\fft{\rho^2 d\theta^2}{\Delta_\theta}\nn\\
&& + \fft{r^2+a^2}{\Xi_a}\sin^2\theta d\phi^2 +
      \fft{r^2+b^2}{\Xi_b} \cos^2\theta d\psi^2\,,\label{5met}\\
A &=& \fft{\sqrt3 q}{\rho^2}\,
         \Big(\fft{\Delta_\theta\, d t}{\Xi_a\, \Xi_b}
       - \omega\Big)\,,\label{gaugepot}
\eea
%%%%%
where
%%%%%
\bea
\nu &=& b\sin^2\theta d\phi + a\cos^2\theta d\psi\,, \qquad \omega =
a\sin^2\theta \fft{d\phi}{\Xi_a} + b\cos^2\theta
\fft{d\psi}{\Xi_b}\,,\nn\\
\rho^2 &=& r^2 + a^2 \cos^2\theta + b^2 \sin^2\theta, \qquad
\Delta_\theta = 1 - a^2 g^2 \cos^2\theta - b^2 g^2
\sin^2\theta\,,\nn\\
\Delta_r &=& \fft{(r^2+a^2)(r^2+b^2)(1+g^2 r^2) + q^2 +2ab q}{r^2} - 2m
\,,\nn \\
f&=& 2 m \rho^2 - q^2 + 2 a b q g^2 \rho^2, \qquad \Xi_a =1-a^2
g^2\,,\quad \Xi_b = 1-b^2 g^2\,.
\eea
The metric is written in Boyer-Lindquist type coordinates, although we
should emphasise that it is in a non-rotating frame at asymptotic
infinity. For general rotation parameters $a$ and $b$ this metric has
three commuting Killing vectors, namely $\partial_t$,
$\partial_{\phi}$ and $\partial_{\psi}$.
Remarkably, one can check that the determinant of the metric is
independent of the charge parameter $q$ and is thus given by the same
expression as in the uncharged case,
\be
\sqrt{-\textrm{det}g} = \frac{r \rho^2 \sin\theta \cos\theta}{\Xi_a \Xi_b}.
\ee
A tedious calculation allows one to write the inverse metric as:
\bea
\rho^2 \, g^{t t} &=& - \frac{ (a^2 + b^2) (2 m r^2 - q^2)}{r^2 \Delta_r}
  -\frac{ (r^2+ a^2) (r^2 + b^2) [ r^2(1 - g^2 (a^2 + b^2)) - a^2
  b^2 g^2] }{ r^2 \Delta_r } \nonumber
\\ &-& \frac{2 m a^2 b^2}{r^2 \Delta_r }- \frac{ 2 a b q r^2 }{ r^2
  \Delta_r } - \frac{ a^2 \cos^2 \theta \, \Xi_a + b^2 \sin^2 \theta
  \, \Xi_b}{ \Delta_{\theta} } \nonumber
\\ \rho^2 \, g^{t \phi} &=&  \frac{ a q^2 - [2 m a + b q (1 + a^2 g^2)]
 (r^2 + b^2)}{ r^2 \Delta_r } \nonumber \\
\rho^2 \, g^{t \psi} &=&  \frac{ b q^2 - [2 m b + a q ( 1 + b^2 g^2)]
  (r^2+ a^2)}{ r^2 \Delta_r } \nonumber \\
\rho^2 \, g^{\phi \phi} &=&  \frac{ a^2 g^2 q^2}{ r^2 \Delta_r } +
\frac{\Xi_a}{\sin^2 \theta} + \frac{\Xi_a}{ r^2 \Delta_r } (1 + g^2
r^2) (r^2+ b^2) (b^2- a^2)   \nonumber \\
&-& \frac{2 m}{ r^2 \Delta_r } (a^2 g^2 r^2
+ b^2)
- \frac{2 a b q}{ \Xi_b \, r^2 \Delta_r } (\Xi_b \, g^2 (r^2 - a^2) -
  b^4 g^4 +1) \nonumber \\
\rho^2 \, g^{\psi \psi} &=&  \frac{ b^2 g^2 q^2}{ r^2 \Delta_r } +
\frac{\Xi_b}{\cos^2 \theta} + \frac{\Xi_b}{ r^2 \Delta_r } (1 + g^2
r^2) (r^2+ a^2) (a^2- b^2)   \nonumber \\
&-& \frac{2 m}{ r^2 \Delta_r } (b^2 g^2 r^2
+ a^2) - \frac{2 a b q}{ \Xi_a \, r^2 \Delta_r } (\Xi_a \, g^2 (r^2 -
b^2) - a^4 g^4 +1) \nonumber \\
\rho^2 \, g^{\phi \psi} &=&  \frac{ a b g^2 q^2 - (1+ g^2 r^2) (2 m a b
  + (a^2 + b^2) q) }{ r^2 \Delta_r } \nonumber \\
\rho^2 g^{\theta \theta} &=& \Delta_{\theta}, \qquad \rho^2 g^{r r}
=\Delta_{r}.
\eea
It is an important fact, that we will use shortly, is that the component
functions $\rho^2 g^{\mu\nu}$ are additively separable as functions of
$r$ and $\theta$. \\

\par
\noindent
The Hamiltonian describing the motion of free uncharged particles in the
background metric $g_{\mu\nu}$ is simply
$H=\frac{1}{2}g^{\mu\nu}p_{\mu}p_{\nu}$. The
corresponding Hamilton-Jacobi equation is then
\be
\frac{\partial S}{\partial\tau} + \frac{1}{2}g^{\mu\nu}
\frac{\partial S}{\partial x^{\mu}}
\frac{\partial S}{\partial x^{\nu}}=0
\ee
where $S$ is Hamilton's principal function and $\tau$ is the parameter
along the worldline of the particle. Due to the presence of the
isometries one may immediately separate out the dependence on
$t,\phi,\psi$ leaving
\be
S= \frac{1}{2}M^2 \tau -E t +L_1\phi +L_2 \psi +F(r,\theta),
\ee
where $M^2$, $E$ and $L_i$ are constants. Remarkably, it turns out that
$S$ is completely separable so $F(r,\theta)=
S_r(r)+S_{\theta}(\theta)$. The proof of this simply relies on the
non-trivial fact that $\rho^2 g^{\mu \nu}$ is
  additively separable as a function of $r$ and $\theta$. This implies that the HJ equation is separable after multiplying it
  through by $\rho^2$. The $\theta$ dependent part of the HJ equation
  is
\bea
\Delta_{\theta} \left( \frac{d S_{\theta}}{d\theta} \right)^2 +
\frac{L_1^2 \Xi_a}{\sin^2\theta} + \frac{L_2^2 \Xi_b }{\cos^2\theta}
  - \frac{E^2}{\Delta_{\theta}}(a^2\Xi_a \cos^2\theta+ b^2 \Xi_b
  \sin^2\theta) + M^2 (a^2\cos^2\theta + b^2 \sin^2\theta) = K
\eea
whilst the $r$ dependent part is \bea \Delta_r \left( \frac{d
S_{r}}{d r} \right)^2 + V(r; E,L_i,M)=-K \eea where $K$ is the
separation constant, and we have defined an ``effective'' potential
$V$ which is a complicated function of $r$; as we shall not use it
directly, we
shall not display it for the sake of brevity. From the $\theta$ equation one may easily read
off a Killing tensor for the spacetime using $K=K^{\mu\nu}
p_{\mu}p_{\nu}$ and $g^{\mu\nu}p_{\mu}p_{\nu} =-M^2$. This gives
\bea
K^{\mu\nu} &=& -g^{\mu\nu}(a^2\cos^2\theta + b^2 \sin^2\theta) -
\frac{1}{\Delta_{\theta}}(a^2\Xi_a \cos^2\theta+ b^2 \Xi_b
  \sin^2\theta) \delta^{\mu}_t \delta^{\nu}_t  \nn \\ &+&
  \frac{\Xi_a}{\sin^2\theta} \delta^{\mu}_{\phi} \delta^{\nu}_{\phi}+
  \frac{\Xi_b }{\cos^2\theta} \delta^{\mu}_{\psi} \delta^{\nu}_{\psi} +
  \Delta_{\theta}\delta^{\mu}_{\theta} \delta^{\nu}_{\theta}.
\eea
Note that this has a smooth limit as $g \to 0$ and coincides with the
Killing tensor found in~\cite{KL}, up to terms which are outer products
of the Killing vectors. In contrast to~\cite{KL}, here it was unnecessary
to add outer products of Killing vectors to the Killing tensor in
order to obtain a smooth limit. This is presumably related to the fact
that we are in a non-rotating frame at infinity, whereas the metric
in~\cite{KL} was in a rotating frame. It is a curious result that the
Killing tensor does not depend explicitly on the charge, although this does also
occur for the four dimensional Kerr-Newman solution. Moreover, as we will discuss
shortly, there exist supersymmetric solutions with $a \neq b$. Such black holes thus possess an \emph{irreducible}
Killing tensor, as do the supersymmetric Kerr-Newman-AdS black
holes in four dimensions~\cite{perry}.
We should note
that from the Hamiltonian point of view the functions
$H,K,p_t,p_{\phi},p_{\psi}$ are in involution thus establishing
Liouville integrability.
The general solution to geodesic motion can easily be deduced from
the generating function $S$ by differentiating with respect to
$K,M^2,E,L_i$ respectively.\\

\par
\noindent
As in the uncharged case, the additive separability of $\rho^2 g^{\mu
  \nu}$ allows for separable solutions to the Klein-Gordon equation
  which governs quantum field theory of massive, spinless particles on
  this background. Writing the KG equation as
\begin{equation}
\frac{1}{\sqrt{-\textrm{det}g}} \, \partial_{\mu} ( \sqrt{-
  \textrm{det} g} \, g^{\mu \nu} \, \partial_{\nu} \Phi ) = M^2 \Phi,
\end{equation}
and taking the following standard ansatz $\Phi = e^{-i \omega t} e^{i
  \alpha \phi} e^{i \beta \psi} R(r) \Theta( \theta)$, renders the KG
  equation separable. The details of this are rather similar to the
  uncharged Kerr-(A)dS~\cite{KL}.
By making the change of variable $z=\sin^2 \theta $, the $\theta$ equation
can be rewritten as:
\begin{equation}
\frac{d^2 \Theta}{d z^2} + \Bigg( \frac{1}{z} + \frac{1}{z-1} +
\frac{1}{z-d}  \Bigg) \frac{ d \Theta}{d z} + \Bigg[ \frac{ \omega^2 (
    a^2 \Xi_a + z(b^2 \Xi_b - a^2 \Xi_a))}{4 z (1-z) \Delta_z^2} -
  \frac{1}{4 z (1-z) \Delta_z} \Bigg(  \frac{\alpha^2 \Xi_a}{ z } +
  \frac{\beta^2 \Xi_b}{ 1-z } \Bigg) \nonumber
\end{equation}
\begin{equation}
+\frac{M^2}{4 g^2 z ( 1-z )} - \frac{k'}{4 z ( 1-z ) \Delta_z}
\Bigg] \Theta = 0,
\end{equation}
where $d = \Xi_a / (g^2 (b^2-a^2))$, $\Delta_z = \Xi_a + g^2 z (a^2 -
b^2)$ and $k' = k + M^2/g^2$ with $k$ being the separation
constant. This equation has four regular singular points and happens
to be in the form of Heun's equation. The special case
$a=b$ simplifies this equation and the solutions are Jacobi
polynomials.\\

\par
\noindent
Having discussed the separability of the Klein-Gordon equation, the
next thing to consider is the Dirac equation on this background. We
find that the Dirac equation separates in the special case of equal
rotation parameters, $a=b$, and can be written as
\bea \label{eq}
(D_r +D_{\theta'}) \Psi = 0,
\eea
where $D_r$ and $D_{\theta'}$ are linear differential operators
depending only on $r$ and $\theta'$ respectively, once the following ansatz has been
made:
\bea
\Psi = e^{-i\omega t} e^{im_1\phi'} e^{im_2 \psi'} \chi(r,\theta').
\eea The angular coordinates $(\theta',\phi',\psi')$ are Euler
angles following the notation of~\cite{KL2}.
This then admits solutions which are separable in the sense that
\bea
\chi(r,\theta' ) = \left(
\begin{array}{c} R_1(r)S_+(\theta') \\ R_2(r) S_-(\theta') \\ R_3(r)
S_+(\theta') \\ R_4(r) S_-(\theta') \end{array} \right),
\eea
where the radial functions form a complicated, coupled system and the functions $S_{\pm}$ are eigenfunctions of the differential
operators
\bea
 \partial^2_{\theta'} + \cot \theta' \partial_{\theta'} -
  \frac{1}{2\sin^2 \theta'} \mp \frac{i(m_1 \cos \theta' - m_2)}{\sin^2
    \theta'} + \frac{\cot^2 \theta'}{4} + \frac{(m_1 - m_2 \cos
    \theta')^2}{ \sin^2 \theta'}.
\eea
\noindent
In four dimensions the separability of the Dirac equation leads to the
construction of an operator that commutes with the Dirac operator, and is
intimately related to the existence of a Yano tensor for the
spacetime~\cite{CaMc}.
Remarkably the four dimensional Kerr-Newman Killing tensor admits a
decomposition in terms of a Yano tensor, arising from the fact that there is a nontrivial supersymmetry on the worldline
of a spinning particle~\cite{GiRiHo}. It is not hard to show that the five dimensional Schwarzschild's Killing tensor
$K$ does not admit a Yano tensor, and hence this suggests that the
full black hole we have been considering does not either.
However, one actually should try to construct an operator that commutes with
the Dirac operator. One expects this to exist due to the presence of
an extra (separation) constant of the system. In four dimensions this is readily achieved, but
seems to rely crucially on the existence of Weyl spinors, and we have been unable to find such an operator in the
five dimensional case. \\

\par
\noindent
Finally, we will briefly discuss the near horizon geometry of the
supersymmetric limit of the black hole. As discussed in \cite{CCLP},
the metric given in equation (\ref{5met}) can be rewritten as:
\bea
d s^2 = - \frac{\Delta_r \Delta_{\theta} r^2 \sin^2 2\theta}{4 (\Xi_a
  \Xi_b)^2 B_{\phi} B_{\psi}} d t^2 + \rho^2 \Bigg( \frac{ d
  r^2}{\Delta_r} + \frac{ d \theta^2}{\Delta_{\theta}} \Bigg) +
B_{\psi} ( d \psi + \nu_1 d \phi + \nu_2 d t)^2 + B_{\phi} (d \phi +
\nu_3 d t)^2, \label{nhg}
\eea
where
\bea
B_{\psi} = g_{\psi \psi}, \quad B_{\phi} = g_{\phi \phi} - \frac{g_{\phi
\psi}^2}{g_{\psi \psi}}, \quad \nu_1 = \frac{g_{\phi \psi}}{g_{\psi
    \psi}}, \quad \nu_2 = \frac{g_{t \psi}}{g_{\psi \psi}}, \quad
\textrm{and} \quad \nu_3 = \frac{g_{t \phi} g_{\psi \psi} - g_{\phi
    \psi} g_{t \psi}}{g_{\phi \phi} g_{\psi \psi} - g_{\phi \psi}^2}.
\eea
In the supersymmetric limit, some simplification of the metric occurs
due to the constraints imposed upon the parameters $q$ and $m$, namely
\bea
q= \frac{m}{1 + a g + b g} , \quad \textrm{and} \quad m = \frac{(a
  +b)(1+ a g) (1+ b g) (1 + a g + b g)}{g}.
\eea
With these restrictions in place, we find that at the horizon
$r_0^2 = g^{-1} (a + b + a b g)$, $\nu_3 + g= 0$, $g+g\nu_1 + \nu_2 =
0$ and all the other functions in the metric
are complicated functions of $\theta$ and the rotation
parameters. To investigate the
near horizon geometry of this metric we first need to go to a frame
which is co-rotating with the horizon. This is effected by the
redefinitions $\tilde{t} = t$, $\tilde{\phi} = \phi - g t$, and $\tilde{\psi}
=  \psi - g t$.
Then we set $r - r_0 = \epsilon R$
and $\tilde{t} = T/\epsilon$ and take the limit $\epsilon
\rightarrow 0$.
The near horizon geometry is then
\begin{eqnarray}\label{nh}
ds^2_{\textrm{NH}} &=& \rho^2(\theta) \left(-c_1 R^2 dT^2 +
  c_2\frac{dR^2}{R^2} +
  \frac{d\theta^2}{\Delta_{\theta}} \right ) + B_{\psi}(\theta) ( d
  \tilde{\psi} + \nu_1(\theta) d \tilde{\phi} +f(\theta)RdT)^2
  \nonumber \\
&+& B_{\phi}(\theta) (d\tilde{\phi}+ c_3 R dT)^2
\end{eqnarray}
where in general we denote $F(\theta) \equiv F(r_0,\theta)$. The
function $f(\theta)$ as well as the constants $c_1,c_2,c_3$ are
complicated and rather unenlightening.  The resulting geometry is
similar to the product of $AdS_2$ with a squashed sphere, which
appears to be a generic property of extremal, rotating (possibly
charged) black holes~\cite{H}. A trivial time rescaling $T =
\sqrt{\frac{c_2}{c_1}}\tilde{T}$ puts the $\tilde{T}R$
part of the metric into a form conformal to $AdS_2$ in Poincar\'e
coordinates. Thus, in addition to the obvious isometries generated by
$\frac{\partial}{\partial \tilde{T}}$, $\frac{\partial}{\partial \tilde
  \phi}$, and $\frac{\partial}{\partial \tilde \psi}$, (\ref{nh}) is
also invariant under dilations $\tilde T \rightarrow \alpha \tilde T,
R \rightarrow R/ \alpha$. An obvious question is whether the near
horizon limit has all the symmetries of $AdS_2$. Following~\cite{H},
one might try to introduce global coordinates on the $AdS_2$, in order
to show the near horizon limit has an (analogue) of the global time
translation. This
needs to be accompanied by a corresponding coordinate transformation
for $(\tilde{\psi}, \tilde{\phi})$. We find that this method does not
work in this case, due to the $\theta$ dependence of the metric.

\noindent
Nevertheless, we can show that the near horizon limit has all the
symmetries of $AdS_2$ in the special case $a=b$ as follows. Let us write the near horizon limit in terms of
left-invariant forms on $SU(2)$ as in~\cite{gr}. It is of the form
\begin{eqnarray}
ds^2 = -(R d \tau + j\sigma_{3})^2 + \frac{dR^2}{R^2} +
L^2(\sigma_{1}^2 + \sigma_{2}^2) + \lambda^2 \sigma_{3}^2
\end{eqnarray} where $j,\lambda,L$ are constants related to the horizon
radius and the cosmological constant. One should note that this metric is a deformation of
the near horizon limit of BMPV as found in~\cite{GMT}. One may easily check that in addition to the time
translation $k = \frac{\partial}{\partial\tau}$ and the dilation
operator $l =-\tau\frac{\partial}{\partial \tau} + R
\frac{\partial}{\partial R}$, there is a third isometry analogous to
the one for pure $AdS_2$:
\begin{equation}
m = \frac{2}{R^2}\left(1 - \frac{j^2}{\lambda^2} \right)
  \partial_{\tau} + 2\tau^2 \partial_{\tau} - 4\tau R \partial_{R} +
  \frac{4j}{R\lambda^2} \partial_{\psi'}.
\end{equation} One may check that these Killing vectors satisfy
\begin{eqnarray}
[l,k] = k, \qquad [l,m] = -m, \qquad [k,m] = -4l.
\end{eqnarray} Furthermore, the gauge field $A$ is regular in the
near-horizon limit and one can easily check that $\pounds_k F =
\pounds_l F = \pounds_m F=0$. Therefore, the algebra of the isometry group of the
near horizon limit which preserves the field strength, in the $a=b$ case, is $\mathfrak{sl}(2,\mathbb{R})
  \times \mathfrak{su}(2) \times \mathfrak{u}(1)$. It would be most
  interesting to see whether the general case retains all the
  symmetries of $AdS_{2}$. Further, an interesting problem is to
  determine the full superalgebra of the near horizon limit, as was
  done for the BMPV case in~\cite{GMT}.

\noindent
The coordinates we are using are not really suitable on the horizon. One should really be using Gaussian
null coordinates adapted to the Killing horizon, which would also allow direct
comparison with the near horizon geometries derived in~\cite{gr} . One
expects the ``parameter'' $\Delta$ used therein to be non-constant for
the metric at hand and thus would fall outside their analysis. \\

\noindent While we have studied certain special symmetries of the general
charged Kerr-AdS black holes, we doubt that the short list
presented here is exhaustive. The existence of
supersymmetric black hole solutions with spherical topology having non-equal
angular momentum in two orthogonal planes seems unique to gauged
supergravity. Given the natural link between supersymmetry and special
geometric structures, it seems likely there are further non-trivial
symmetries of these black holes.
\\
\\
\noindent
PD would like to thank PPARC for financial support. HKK would like
to thank St. John's College for financial support. We would like to
thank Gary Gibbons and Harvey Reall for useful comments.


\begin{thebibliography}{99}

\bibitem{Carter1}
B.~Carter,
``Global Structure Of The Kerr Family Of Gravitational Fields,''
Phys.\ Rev.\  {\bf 174} (1968) 1559.

\bibitem{Carter2}
B.~Carter,
``Hamilton-Jacobi And Schrodinger Separable Solutions Of Einstein's
Equations,''
Commun.\ Math.\ Phys.\  {\bf 10} (1968) 280.

\bibitem{GLPP}
G.~W.~Gibbons, H.~Lu, D.~N.~Page and C.~N.~Pope,
``The general Kerr-de Sitter metrics in all dimensions,''
J.\ Geom.\ Phys.\  {\bf 53} (2005) 49, {\tt hep-th/0404008}.

\bibitem{KL}
H.~K.~Kunduri and J.~Lucietti,
``Integrability and the Kerr-(A)dS black hole in five dimensions,''
Phys.\ Rev.\ D {\bf 71} (2005) 104021, {\tt hep-th/0502124}.

\bibitem{VSP}
M.~Vasudevan, K.~A.~Stevens and D.~N.~Page,
``Separability of the Hamilton-Jacobi and Klein-Gordon equations in  Kerr-de
Sitter metrics,''
Class.\ Quant.\ Grav.\  {\bf 22} (2005) 339, {\tt gr-qc/0405125}.

\bibitem{VS}
M.~Vasudevan and K.~A.~Stevens,
``Integrability of particle motion and scalar field propagation in Kerr-(anti)
de Sitter black hole spacetimes in all dimensions,''
{\tt gr-qc/0507096}.

\bibitem{CLP}
  M.~Cvetic, H.~Lu and C.~N.~Pope,
  ``Charged Kerr-de Sitter black holes in five dimensions,''
  Phys.\ Lett.\ B {\bf 598} (2004) 273, { \tt hep-th/0406196}.

\bibitem{KL2}
  H.~K.~Kunduri and J.~Lucietti,
`Notes on non-extremal, charged, rotating black holes in minimal D = 5 gauged
supergravity,''
Nucl.\ Phys.\ B {\bf 724} (2005) 343
{\tt hep-th/0504158}.

\bibitem{vasu}
M.~Vasudevan,
``Integrability of some charged rotating supergravity black hole solutions in
four and five dimensions,''
{ \tt gr-qc/0507092}.


\bibitem{CCLP}
Z.~W.~Chong, M.~Cvetic, H.~Lu and C.~N.~Pope,
``General non-extremal rotating black holes in minimal five-dimensional gauged
supergravity,'' {\tt hep-th/0506029}

\bibitem{GMT}
J.~P.~Gauntlett, R.~C.~Myers and P.~K.~Townsend,
``Black holes of D = 5 supergravity,
Class.\ Quant.\ Grav.\  {\bf 16} (1999) 1, {\tt hep-th/9810204}

\bibitem{perry}
V.~A.~Kostelecky and M.~J.~Perry,
``Solitonic Black Holes in Gauged N=2 Supergravity,''
Phys.\ Lett.\ B {\bf 371} (1996) 191{\tt hep-th/9512222}.

\bibitem{CaMc}
B.~Carter and R.~G.~Mclenaghan,
``Generalized Total Angular Momentum Operator For The Dirac Equation In Curved
Space-Time,''
Phys.\ Rev.\ D {\bf 19} (1979) 1093.

\bibitem{GiRiHo}
 G.~W.~Gibbons, R.~H.~Rietdijk and J.~W.~van Holten,
  ``SUSY in the sky,''
  Nucl.\ Phys.\ B {\bf 404} (1993) 42, {\tt hep-th/9303112}.

\bibitem{H}
J.~M.~Bardeen and G.~T.~Horowitz,
`The extreme Kerr throat geometry: A vacuum analog of AdS(2) x S(2),''
Phys.\ Rev.\ D {\bf 60} (1999) 104030, {\tt hep-th/9905099}.

\bibitem{gr}
J.~B.~Gutowski and H.~S.~Reall,
``Supersymmetric AdS(5) black holes,''
JHEP {\bf 0402} (2004) 006, {\tt hep-th/0401042}.


\end{thebibliography}
\end{document}